\documentclass[twocolumn,amsmath,amssymb,pra,showpacs]{revtex4}
\usepackage{graphicx}

\begin{document}

\title{Resonant control of cold-atom transport through two optical lattices with a constant relative speed}
\author{M.T. Greenaway$^1$,  A.G. Balanov$^{2}$, and T.M. Fromhold$^1$}
\affiliation{$^1$School of Physics and Astronomy, University of Nottingham, Nottingham NG7 2RD, United Kingdom \\
$^2$Department of Physics, Loughborough University, Leicestershire, LE11 3TU, United Kingdom}
\date{\today}

\begin{abstract}
We show theoretically that the dynamics of cold atoms in the lowest energy band of a stationary optical lattice can be transformed and controlled by a second, weaker, periodic potential moving at a constant speed along the axis of the stationary lattice.  The atom trajectories exhibit complex behavior, which depends sensitively on the amplitude and speed of the propagating lattice.  When the speed and amplitude of the moving potential are low, the atoms are dragged through the static lattice and perform drifting orbits with frequencies an order of magnitude higher than that corresponding to the moving potential.  Increasing either the speed or amplitude of the moving lattice induces Bloch-like oscillations within the energy band of the static lattice, which exhibit complex resonances at critical values of the system parameters. In some cases, a very small change in these parameters can reverse the atom's direction of motion.  In order to understand these dynamics we present an analytical model, which describes the key features of the atom transport and also accurately predicts the positions of the resonant features in the atom's phase space.   
The abrupt controllable transitions between dynamical regimes, and the associated set of resonances, provide a mechanism for transporting atoms between precise locations in a lattice: as required for using cold atoms to simulate condensed matter or as a stepping stone to quantum information processing.  The system also provides a direct quantum simulator of acoustic waves propagating through semiconductor nanostructures in sound analogs of the optical laser (SASER). 

\end{abstract}

\maketitle{}

\section{Introduction}

Cold atoms in optical lattices created by two or more counter-propagating laser beams provide a clean, flexible, physical environment for studying quantum particles in periodic potentials. Key advantages of such systems include the tunability of the lattice, low dissipation, and the ability to directly control and measure the atoms' internal state.  When the atoms are cold enough for the de Broglie wavelength, corresponding to the atoms' center-of-mass motion, to span several wells of the lattice potential, the associated energy eigenvalues form bands of allowed energies separated by band gaps. The dynamics of atoms in energy bands is of particular interest due to the similarities with the behavior of electrons in solids \cite{MOR2006, BON2004, LIG2007, TAR2012, BEC2010, PON2006}.  For example, atoms confined to a single energy band and subject to a constant force have been shown to Bragg reflect and perform many Bloch oscillations \cite{BEN1996,MOR2001,SCO2003,GUS2008,SAL2009,SAL2008}: a phenomenon well known, but harder to realise, in solid state physics \cite{BLO1929,ASH1979}.  

Recently, there has been increased experimental interest in nonlinear phenomena relating to the transport of atoms in optical lattices. In particular, it has been shown that moving optical lattices can transport cold atoms along distances as far as 20 cm \cite{SCHM2006}. In addition, unusual band structure properties have been revealed by experiments with a moving optical lattice \cite{BRO2005}, critical velocities in such systems have been identified \cite{AHA2009}, the stability of superfluid currents in systems of ultracold bosons have been studied \cite{MUN2007}, and the phenomenon of stochastic resonances has been predicted to play an important role in transport through dissipative optical lattices \cite{SCH2002}. For single-atom transport, both lensing \cite{FAL2003} and the dynamics and control \cite{FAL2004, DAB2006, EIE2003, CHO2004, CHA2006, ARL2011} of BECs have been observed and studied.

In this paper, we propose a method to control precisely the dynamics of atoms moving through a {\it stationary} optical lattice by introducing an additional {\it moving} periodic optical potential. We demonstrate that the combined effect of the two lattices induces multiple sharp resonances, which can be exploited to tailor the atomic motion.  Also, this system has a direct counterpart in semiconductor physics. Specifically, it models the electron dynamics induced by an acoustic wave propagating through semiconductor nanostructures \cite{GRE2010}: a topic of growing interest due to the recent development of coherent sources of phonons (SASERs) \cite{WAL2009, BEA2010, BEA2011}, which are analogs of the optical laser.

The paper has the following structure. In Section \ref{sec:semiclass}, we introduce the semiclassical  equations of motion for an atom in a stationary
optical lattice driven by a propagating optical potential. Analysis of the equations reveals the distinct dynamical regimes of this system and the abrupt transitions between them. In Section \ref{sec:lat_speed}, we study how the speed of the propagating potential affects the average velocity of atoms in the stationary lattice. Section \ref{sec:init_pos} explores how the atom's initial position affects transitions between the various dynamical regimes. In Section \ref{sec:quantum}, we compare our semiclassical results with full quantum calculations.  Finally, in Section \ref{sec:conc}, we draw conclusions and propose experiments to study the transport phenomena that we have identified.

\begin{figure}%f1
\centering
 \includegraphics*[width=.8\linewidth]{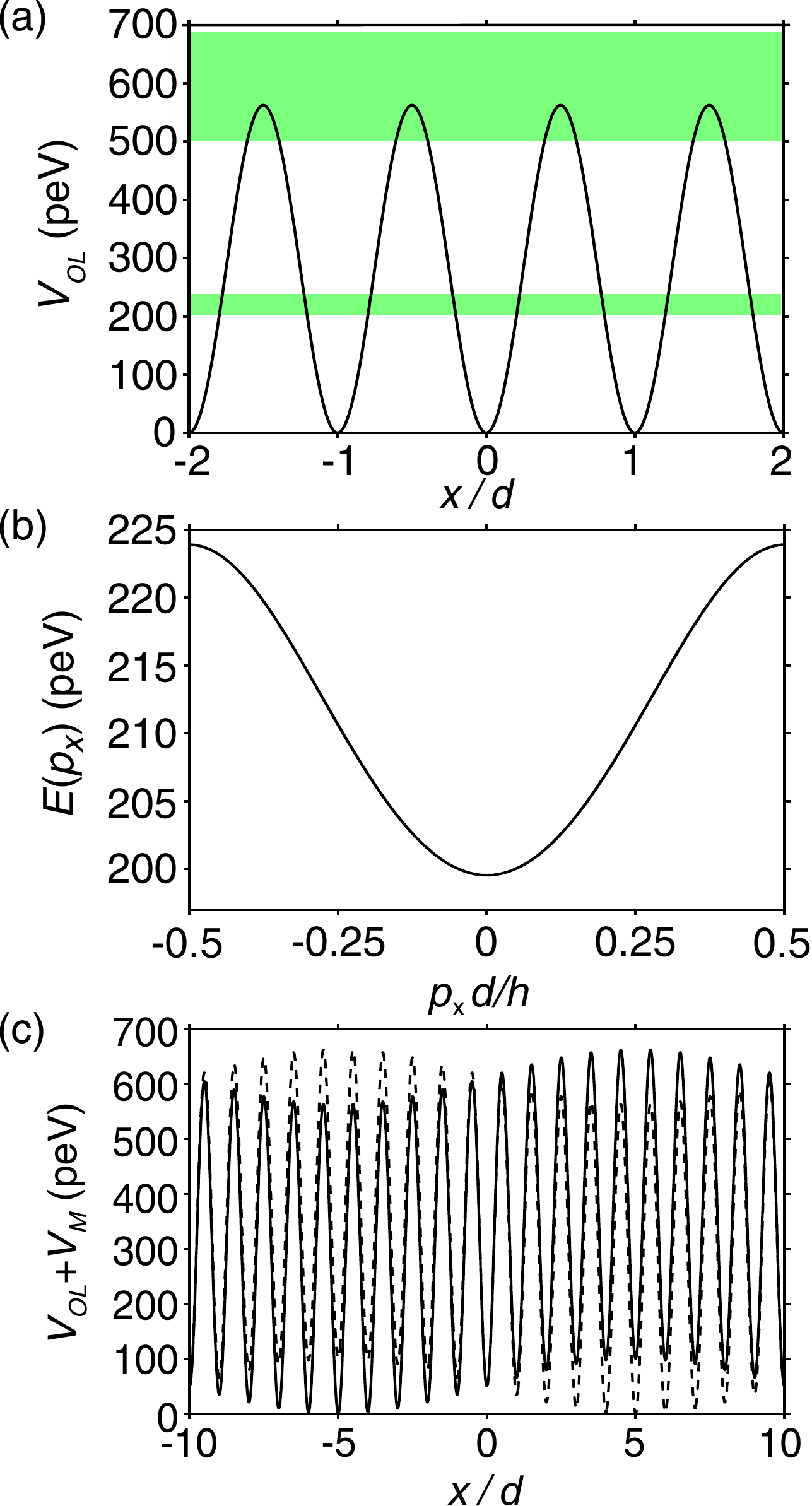}
   \caption{(a) solid curve shows the potential energy, $V_{OL}$, of a $^{23}$Na atom in the stationary optical lattice.  Shaded regions show energy ranges of first and second energy bands.  (b) $E(p_x)$ curve calculated for the first energy band. (c) Solid curve: total potential, $V_{OL}+V_M$, of the atom at $t=0$. Dashed curve: total potential after half a temporal period of $V_M$ (i.e. at $t=\pi/\omega_M$).  
 \label{fig:1}}
\end{figure}

\section{Semiclassical atom trajectories}
\label{sec:semiclass}

We consider a cloud of cold, non-interacting, sodium atoms \cite{GUS2008,SCO2002,DAV1995,WIL1996, BHA1999} initially in the lowest energy band of a 1D
\emph{stationary} optical lattice (SOL), taking the potential energy profile of each atom to be $V_{OL}(x)=V_0 \sin^2 \left(\pi x / d \right)$ [Fig.
\ref{fig:1}(a)].  The lattice period $d=294.5$ nm and its depth is $V_0=563$ peV $=5.5E_R$, where $E_R=103$ peV is the recoil energy.  The energy versus crystal momentum dispersion relation for the lowest energy band is, to good approximation, $E(p_x)=E_0 + \Delta\left[1-\cos(p_xd/\hbar)\right]/2$ [Fig. \ref{fig:1}(b)], where $E_0\approx200$ peV and $\Delta=24.4$ peV is the band width.  To this system we apply an additional \emph{moving} optical potential (MOP), which propagates along the $x$ axis at velocity, $v_M$, creating a position and time ($t$) dependent potential energy field, $V_M(x,t) = U_M\left[1- \sin \left( k_M x - \omega_M t \right)\right]/2$ [Fig. \ref{fig:1}(c)], where the wave number $k_M = 2 \pi / \lambda_M$, $\lambda_M$ is the wavelength, and $\omega_M=k_M v_M$ is the angular frequency.  Such a field can be generated by counter-propagating laser beams whose frequencies are slightly detuned by $\delta \omega$. The velocity of the MOP is then $v_M = \delta \omega / 2 k_M$ \cite{SCH2002,SCHM2006,BRO2005}.  

The semiclassical Hamiltonian for an atom in the lowest energy band of the stationary optical lattice, and subject to the moving potential, is $H(x,p_x,t)=E(p_x) + V_M(x,t)$.  The corresponding Hamilton's equations of motion are

\begin{eqnarray}
  \frac{\partial p_x}{\partial t} = \frac{U_M}{2} k_M \cos \left[k_M (x+x_0) - \omega_M t \right] \label{eq:dpxdt}\\
  v_x = \frac{\partial x}{\partial t} = \frac{\Delta d}{2 \hbar} \sin \left( \frac{p_x d}{\hbar}\right). \label{eq:dxdt}
\end{eqnarray}

We solve Eqs. (\ref{eq:dpxdt}) and (\ref{eq:dxdt}) numerically to calculate the trajectories starting from rest with $x_0=x(t=0)=0$ and $v_x(t=0)=0$. To characterise the transport of an atom through the SOL, we calculate the time averaged velocity, $\langle v_x \rangle_t$, over a period of $0.25$ s. 
 
\begin{figure}%f1
\centering
 \includegraphics*[width=.8\linewidth]{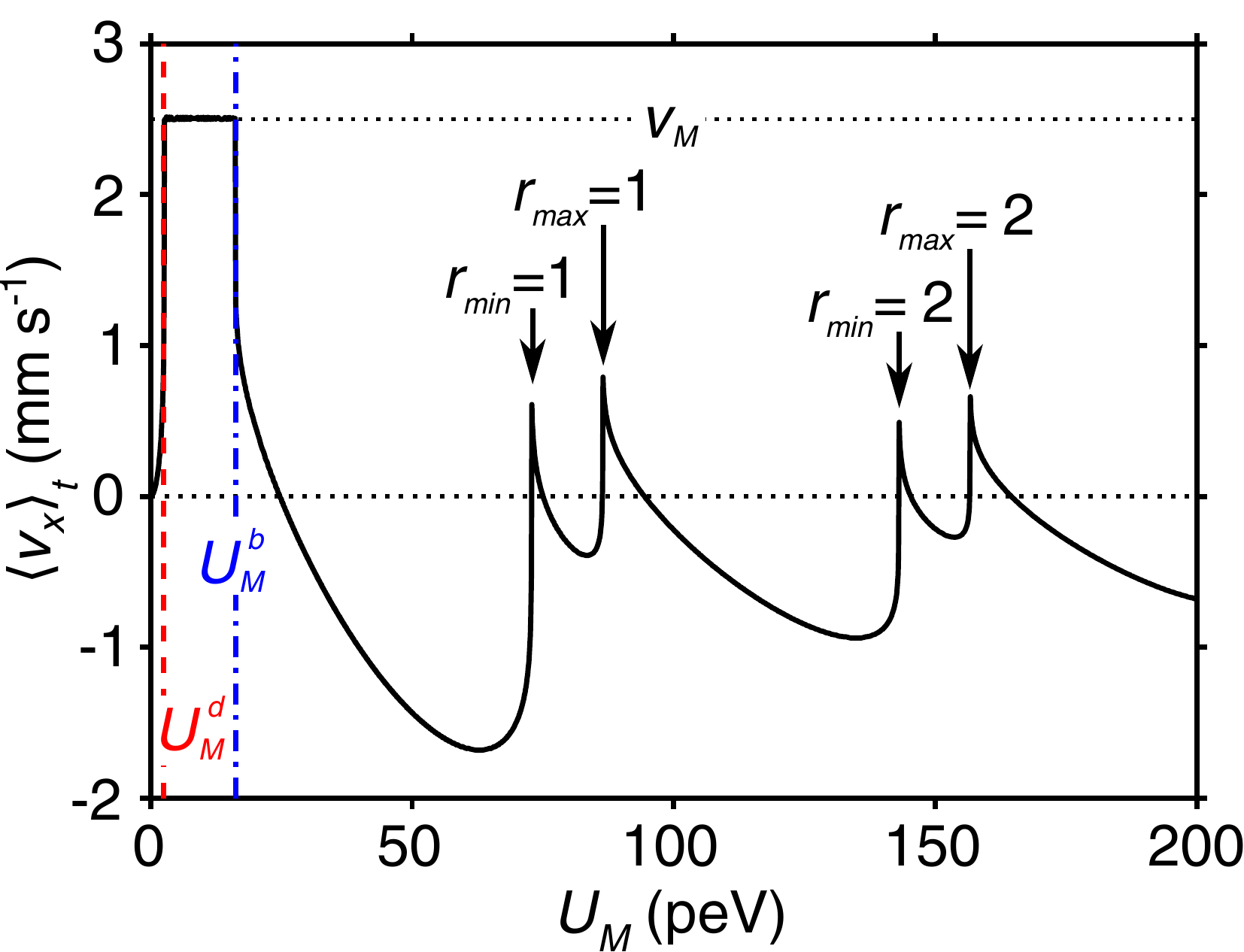}
   \caption{Solid curve shows $\langle v_x \rangle_t$ versus the moving optical potential amplitude, $U_M$.  Vertical (red) dashed line marks the onset of atom dragging (labelled $U_M^d$). Dot-dashed vertical line (blue) marks the onset of Bloch oscillations (labelled $U_M^b$).  Arrows highlight resonant peaks in $\langle v_x \rangle_t$, explained in the text.  For guidance, the upper dotted line shows $\langle v_x \rangle_t=v_M$ and the lower dotted line shows $\langle v_x \rangle_t=0$.  
 \label{fig:2}}
\end{figure}

The solid curve in Fig. \ref{fig:2} shows $\langle v_x \rangle_t$ calculated as a function of $U_M$ for $\lambda_M=20d$ and $v_M=2.5$ mm s$^{-1}$.  The curve reveals that for very low $U_M$, $\langle v_x \rangle_t$ increases exponentially from $0$ with increasing $U_M$, until $U_M$ reaches a critical value, $U_M^d\approx2.5$ peV = $2.43 \times 10^{-2}E_R$ (red vertical dashed line),  at which point $\langle v_x \rangle_t=v_M$ (marked by upper horizontal dotted line).  As $U_M$ increases beyond $U_M^d$, $\langle v_x \rangle_t$ remains pinned at $v_M$ until $U_M$ reaches a second critical value, $U_M^b \approx 16$ peV = $0.156E_R$, (blue vertical dot-dashed line) where the average velocity begins to decrease abruptly with increasing $U_M$.   Thereafter, increasing $U_M$ gives rise to a series of resonant peaks (arrowed in Fig. \ref{fig:2}).

\begin{figure}%f1
\centering
\includegraphics*[width=1.\linewidth]{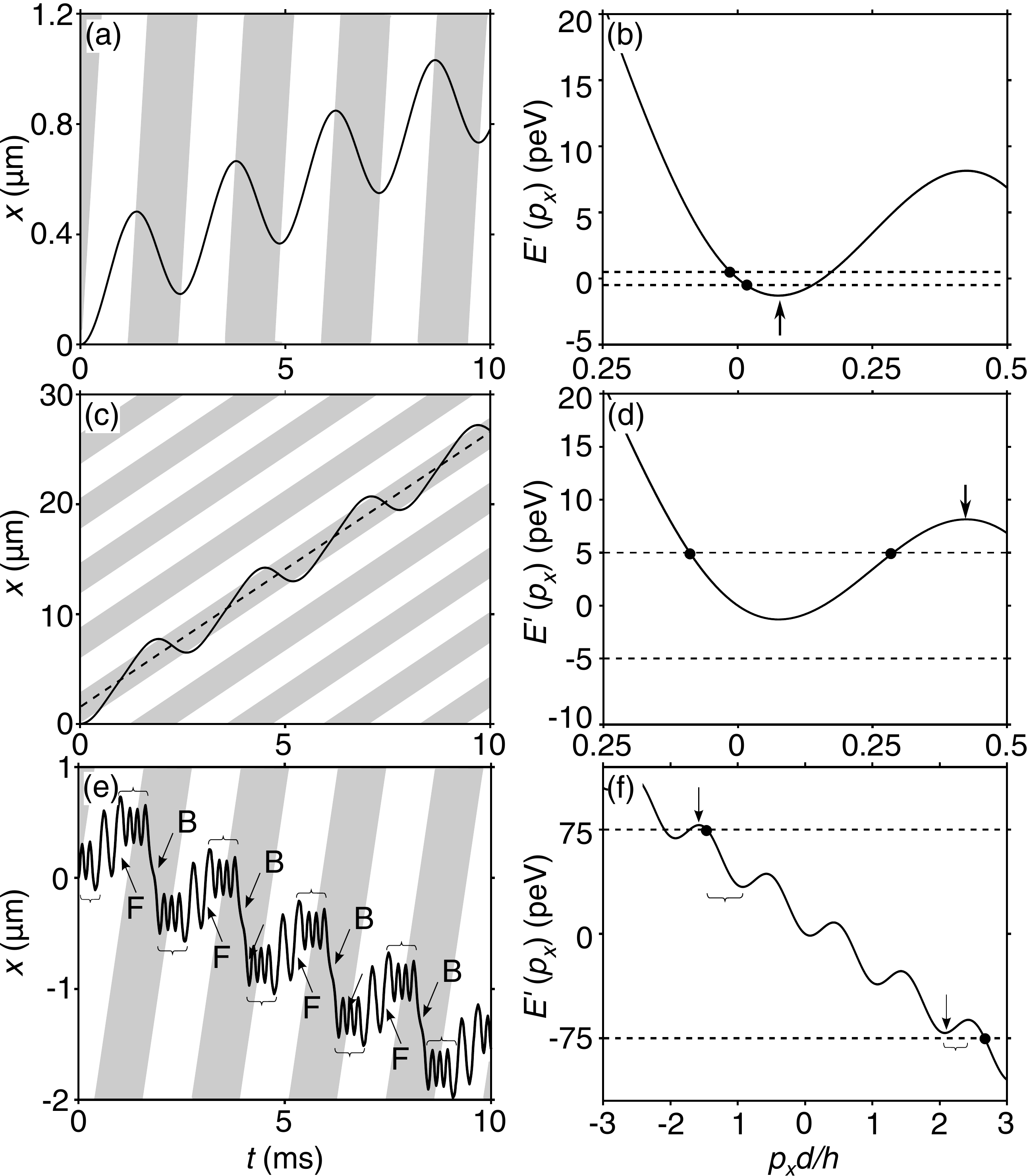}
\caption{$x(t)$ [(a), (c) and (e)] and $E'(p_x)$ [(b), (d) and (f)] curves calculated for $U_M=1$ peV = $9.74 \times 10^{-3}E_R$ [(a) and (b)], $10$ peV = $9.74 \times 10^{-2}E_R$ [(c) and (d)], and $150$ peV = $1.46E_R$ [(e) and (f)].  In (a), (c) and (e), $V_M(x, t)$ is $<U_M/2$ [$>U_M/2$], within the white [gray] regions. In (b), (d) and (f), horizontal dashed lines show $\pm U_M/2$ and filled circles mark when $E'(p_x) = \pm U_M/2$ and, therefore, the limits of the orbits in $p_x$.  Arrows, labels, and brackets are discussed in the text and Appendices \ref{app:1} and \ref{app:2}.
\label{fig:3}}
\end{figure}

To understand the form of the $\langle v_x \rangle_t$ versus $U_M$ curve, in Fig. \ref{fig:3} we show atom trajectories for three values of $U_M$ corresponding to three distinct dynamical regimes.  

Figure \ref{fig:3}(a) shows $x$ versus $t$ for a trajectory %(starting from rest $x(t=0)=p_x(t=0)=0$) 
within the low field regime when $U_M=1$ peV = $9.74 \times 10^{-3}E_R$ ($<U_M^d$, which is marked by red dashed line in Fig. \ref{fig:2}).  In this case, the atom orbit comprises a uniform drift, with mean velocity $\approx75$ $\mu$m s$^{-1}$, and superimposed periodic oscillations.  The electron velocity $v_x>0$ when $V_M(x,t)<U_M/2$ [white regions in Fig. \ref{fig:3}(a)] and $v_x<0$ when $V_M(x,t)>U_M/2$ [gray regions in Fig. \ref{fig:3}(a)].  We explain the form of this trajectory by considering atom motion in the rest frame of the MOP, in which the atom's position is $x'(t)=x(t)-v_Mt$.  In this frame, the semiclassical Hamiltonian is $H'\left(x',p_x\right)=E'(p_x)+V_M(x')$, where $E'(p_x)=E(p_x)-v_Mp_x$ is the effective dispersion relation and the potential, $V_M(x')= U_M\left[1 - \sin \left( k_M x' \right)\right]/2$, which does not explicitly depend on time.   The transformed semiclassical equations of motion are

\begin{eqnarray}
  \frac{\partial p_x}{\partial t} = \frac{U_M}{2} k_M \cos \left[k_M (x'+x_0)\right] \label{eq:dpxdtvdash}\\
  v_x' = v_x - v_M= \frac{\partial x'}{\partial t} = \frac{\Delta d}{2 \hbar} \sin \left( \frac{p_x d}{\hbar}\right)-v_M. \label{eq:dxdtvdash}
\end{eqnarray}
 
Since $H'$ is time-independent, it is a constant of motion but does {\it not} equal the total system energy, $H$. 
For initial conditions $x(t=0)=x_0$ and $p_x(t=0)=p_0$ we obtain

\begin{equation}
 H'(x',p_x)=\frac{U_M}{2}\left[1-\sin\left(k_Mx_0\right)\right].
\label{eq:hamallt}
\end{equation}

%\noindent
Therefore, if the initial position of the atom is $x_0=0$, the constant of motion is $H'=U_M/2$, which means that the kinetic energy in the moving frame $E'(p_x)=H'(x',p_x)-V_M(x')=U_M/2-V_M(x')$ can only take values between $\pm U_M/2$, marked by the horizontal dashed lines in Fig. \ref{fig:3}(b) for $U_M=1$ peV.  This figure reveals that if $p_0=0$, the atom can only access a very limited, almost linear, region of $E'(p_x)$ near $p_x=0$ (between the horizontal dashed lines), where $v_x'=\partial E'(p_x) / \partial p_x \approx -\nu$, with $\nu$ being a positive constant. Consequently, in the rest frame, the mean velocity of the atom, $\langle v_x \rangle_t \approx v_M-\nu$, is smaller than $v_M$.  This can be seen from Fig. \ref{fig:3}(a), where the mean (drift) velocity of the atom orbit, is less than the slope of the grey and white stripes, which equals $v_M$. As the atom traverses successive minima and maxima [white and gray stripes in Fig. \ref{fig:3}(a)] of the moving lattice, the corresponding force changes sign, thereby causing the atom to oscillate between the extremal $p_x$ values marked by the circles in Fig. \ref{fig:3}(b) and, in real space, oscillate around the mean drift [Fig. \ref{fig:3}(a)]. As $U_M$ increases [i.e. the horizontal dashed lines in Fig. \ref{fig:3}(b) move further apart], the atom can access more of the $E'(p_x)$ dispersion curve.  As a result, $\nu$ becomes smaller and $\langle v_x \rangle_t$ increases towards $v_M$, as shown by the section of the black curve in Fig. \ref{fig:2} to the left of the vertical red dashed line. Atom motion in this `Linear Dispersion' (LD) regime is described in detail in Appendix \ref{app:1}.

The LD regime persists with increasing $U_M$ until the value of $-U_M/2$ [lower dashed line in Fig. \ref{fig:3}(b)] falls below the value of the local minimum in $E'(p_x)$ [arrowed in Fig. \ref{fig:3}(b)]. Using the fact that in this regime $|p_x|\ll \pi\hbar/d$, we find that the $p_x$ value corresponding to the local minimum of  $E'(p_x)$ is $p_x\approx 2\hbar^{2} v_M/ \Delta d^{2}$. Substituting this value into the expression for $E'(p_x)$, we obtain the following estimate for the critical amplitude of the moving lattice, $U_M^d$, above which the atom can reach and traverse the local minimum in $E'(p_x)$:
 
\begin{align} 
 U_M^d = \frac{1}{1 + \sin k_M x_0} & \times \label{eq:Umd} \\ \nonumber
 \left[\frac{4 \hbar^{2} v_M^2} {\Delta d^{2}} \right.&\left. - \Delta\left(1-\cos\left(\frac{2\hbar v_M}{\Delta d}\right)\right)\right].
\end{align}

%\noindent
For the parameters considered, Eq. (\ref{eq:Umd}) gives $U_M^d\approx2.5$ peV = $2.43 \times 10^{-2}E_R$, which compares well with the transition to the region where $\langle v_x \rangle_t=v_M$ in Fig. \ref{fig:2} (i.e. to the right of the vertical red dashed line).  

The solid curve in Fig. \ref{fig:3}(c) shows the trajectory of the atom when $U_M=10$ peV = $9.74 \times 10^{-2} E_R > U_M^d$.   We find regular, almost periodic, oscillations
superimposed on a uniform drift, indicated by the dashed line in Fig. \ref{fig:3}(c). The slope of the line is $v_M$, indicating that the atom is
dragged through the SOL by the MOP.  This is confirmed by the fact that the trajectory is confined to a single minimum in the MOP potential where
$V_M(x,t)<0$ [gray region in Fig. \ref{fig:3}(c)], which traps the atom and transports it through the SOL.  Fig. \ref{fig:3}(d) shows that the atom is able to oscillate in a parabolic region of $E'(p_x)$ between the two $p_x$ values (filled circles) where $E'(p_x)=U_M/2$.  Since the atom remains in the almost parabolic region of $E'(p_x)$, $x'(t)$ is an almost harmonic function of $t$.  Therefore we can approximate the trajectory in Fig. \ref{fig:3}(c) by the expression $x(t)\approx v_M t+\lambda_M \left[1-\cos\left(\omega_R t\right)\right]/4$, where $\omega_R$ is the frequency for motion to and fro across the potential well \cite{GRE2010}. In this `Wave Dragging' (WD) regime,  the atom is trapped in a single well of the MOP, but as the well moves, it drags the atom through the SOL with $\langle v_x \rangle_t=v_M$ (upper horizontal dotted line in Fig. \ref{fig:2}).

Increasing $U_M$ in the WD regime initially has no qualitative effect on the trajectories, which continue to be dragged and are of the form $x(t)=v_M t + f(t)$. However, as $U_M$ increases, the atom can access increasingly non-parabolic parts of $E'(p_x)$ and thus $f(t)$ becomes less harmonic.  Eventually, though, $U_M$ becomes large enough for the atom to traverse the first local maximum of $E'(p_x)$ [arrowed in Fig. \ref{fig:3}(d)]. Thereafter, the atom can reach the edge of the first Brillouin zone and its trajectory in $p_x$ changes abruptly from closed to open. In this regime, the atom can traverse several minizones, allowing it to Bragg reflect and perform Bloch oscillations. The first local maximum of $E'(p_x)$ occurs when $p_x  \approx \hbar \pi/d - 2 \hbar^{2} v_M / \Delta d^{2}$ \cite{GRE2010}. Substituting this value into $E'(p_x)$ we find the following approximate expression for  the wave amplitude, $U^b_M$, corresponding to the transition between the WD and `Bloch Oscillation' (BO) regimes, i.e. for which the value of $U_M/2$, shown by the upper dashed line in Fig. \ref{fig:3}(d) reaches the local maximum in $E'(p_x)$ \cite{GRE2010}:

\begin{equation}
 U^b_M \approx \frac{2}{1-\sin k_M x_0}\left[\Delta - \frac{\hbar \pi v_M}{d} + \frac{2 \hbar^2 v_M^2}{\Delta d^2}\right].
\label{eq:Umb}
\end{equation}

%\noindent
The vertical blue dot-dashed line in Fig. \ref{fig:2} shows the value of $U^b_M$ obtained from Eq. (\ref{eq:Umb}) for $x_0=0$, which coincides exactly with the abrupt suppression of $\langle v_x \rangle_t$.  

Figure \ref{fig:3}(e) shows the atom's trajectory when $U_M=150$ peV = $1.46E_R > U_M^b$.  In this regime, the atom undergoes fast oscillations [bracketed in Fig \ref{fig:3}(e)], interrupted by jumps in the orbit (arrowed).  Initially, for $t=0$ and $x(t=0)=p_x(t=0)=0$, the driving force on the atom is maximal, and therefore $p_x$ increases rapidly.  As a result, the atom traverses several local maxima and minima in $E'(p_x)$ [between circles in Fig. \ref{fig:3}(f)], Bragg reflecting and performing Bloch oscillations [within leftmost bracketed region of trajectory in Fig. \ref{fig:3}(e)] as it does so.  As $E'(p_x)$ decreases, $V_M$ increases to keep $H'$ constant.  Eventually, $V_M$ attains its maximum value of $U_M/2$ corresponding to the minimum attainable value of $E'(p_x)\approx -U_M/2$ [lower dashed line in Fig. \ref{fig:3}(f)] and maximum attainable value of $p_x$ [right-hand filled circle in Fig. \ref{fig:3}(f)].  Thereafter, $p_x$ starts to decrease, triggering another burst of Bloch oscillations, until $p_x$ reaches its minimum possible value [left-hand filled circle in Fig. \ref{fig:3}(f)].  Subsequently, $p_x$ increases again and the cycle repeats.  The details of the form of this orbit are given in Appendix \ref{app:2}.

The analysis in the appendix shows that the atom's average velocity will exhibit a resonant peak whenever $U_M$ just exceeds a local maximum, or falls below a local minimum, in $E'(p_x)$, thus maximising the time the atom spends in a region of $E'(p_x)$ where $v_x'=dE'(p_x)/dp_x>0$, and so also maximising $\langle v_x \rangle_t=\langle v_x' \rangle_t+v_M$ .  The minima and maxima in the dispersion
curve occur, respectively, at $p_x$ values given by

\begin{equation}
 p_x^{r_{min}} \approx r_{min}\frac{2\pi\hbar}{d} + \frac{2 \hbar^{2} v_M} {\Delta d^{2}},
\label{eq:pxb}
\end{equation}
 
\noindent
and 

\begin{equation}
 p_x^{r_{max}} \approx -(2r_{max}-1)\frac{\pi\hbar}{d} - \frac{2 \hbar^{2} v_M} {\Delta d^{2}},
\label{eq:pxt}
\end{equation}

\noindent 
where $r_{min}$ and $r_{max}$ are integers ($=1,2,3...$) labelling the minima and maxima, respectively, in $E'(p_x)$.  The corresponding critical values of $U_M$, obtained from setting $E'(p_x)=E(p_x)-v_M p_x=\pm U_M/2$, are 

\begin{equation}
U_M^{r_{min}} \approx - 2E\left(p_x^{r_{min}}\right) + 2 v_M  p_x^{r_{min}}, 
\label{eq:UMrb}
\end{equation}

\noindent
and

\begin{equation}
U_M^{r_{max}} \approx 2E\left(p_x^{r_{max}}\right) - 2 v_M  p_x^{r_{max}}.
\label{eq:UMrt}
\end{equation}

%\noindent
The values of $U_M^{max}$ and $U_M^{min}$ for $r_{max}$ and $r_{min}= 1$ and $2$ are shown in Fig. \ref{fig:2} by the labelled arrows and correspond very well with the positions of the resonant peaks in $\langle v_x \rangle_t$.  

\section{Variation of moving lattice speed}
\label{sec:lat_speed}

The parameters of a moving optical potential can be varied over a wide range \cite{SCHM2006,WIL1996}. For example, it has been shown that $v_M$ can be increased to $\sim 50$ mm s$^{-1}$ \cite{SCHM2006} and $U_M$ to $\sim500$ peV \cite{WIL1996}. 

\begin{figure}%f1
 \centering
 \includegraphics*[width=1.\linewidth]{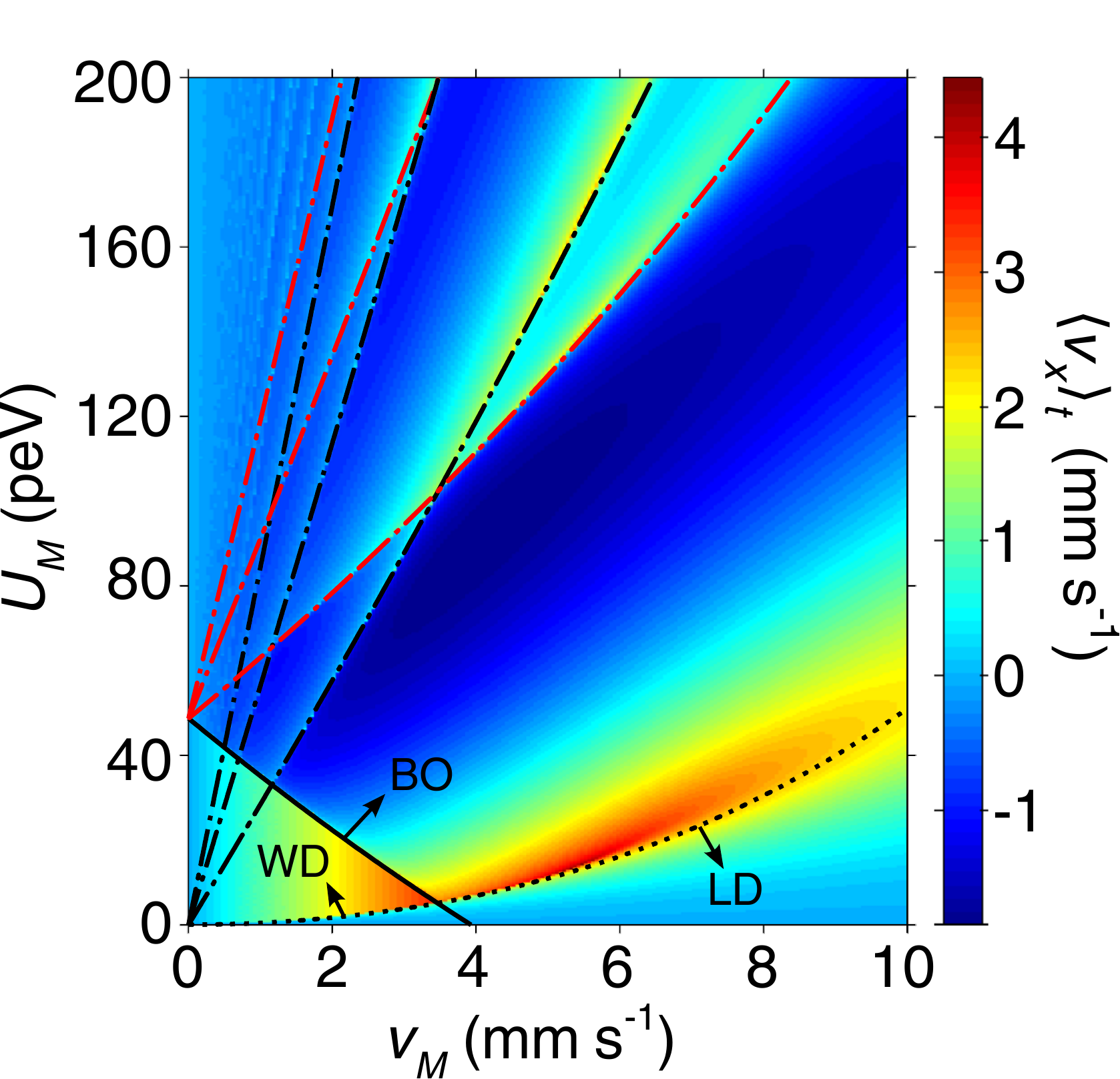}
  \caption{Color map (scale right) showing the variation of $\langle v_x \rangle_t$ with $v_M$ and $U_M$.  Dotted black curve shows $U_M^d$ versus $v_M$, and marks the onset of dragged atom trajectories [Eq. (\ref{eq:Umd})].  Black solid curve shows $U_M^b$ versus $v_M$ [Eq. (\ref{eq:Umb})].  Black dot-dashed curves show $U_M^{r_{min}}$ versus $v_M$ when $r_{min}=1,2$ and $3$ (bottom to top) [Eq. (\ref{eq:UMrb})].  Red dot-dashed curves mark values of $U_M^{r_{max}}$ when $r_{max}=1,2$ and $3$ (bottom to top) [Eq. (\ref{eq:UMrt})].  Labels LD, WD and BO mark regions of `Linear Dispersion', `Wave Dragging' and `Bloch Oscillations'.
\label{fig:4}}
\end{figure}

The color map in Fig. \ref{fig:4} shows the variation of the average velocity of the atom, $\langle v_x \rangle_t$, with both $v_M$ and $U_M$.  The figure reveals a complex patchwork of distinct dynamical regimes.  At the boundaries between these regimes, the average velocity changes abruptly and may even change sign, indicating that the atom's response to the moving wave depends critically on both the amplitude and velocity of that wave. 

The structure of the color map in Fig. \ref{fig:4} can be understood by considering the transitions between the LD, WD and BO regimes discussed in the previous section. The dotted black curve in Fig. \ref{fig:4} shows $U_M^d$ versus $v_M$ [Eq. (\ref{eq:Umd})],
marking the transition between the LD and the WD regimes. For $v_M\lesssim3.5$ mm s$^{-1}$ (i.e. to the left of the intersect between the solid and dotted black curves), increasing $U_M$ across the black dotted curve induces a clear transition from the LD regime, where $\langle v_x \rangle_t \approx 0$
(blue in Fig. \ref{fig:4}), to the WD regime, where $\langle v_x \rangle_t\approx v_M$ (yellow / red in Fig. \ref{fig:4}).  

The black solid curve in Fig. \ref{fig:4} shows $U_M^b$ versus $v_M$ [Eq. (\ref{eq:Umb})] and thus marks the transition from the WD regime into the BO regime.  The region of the color map below the black solid curve and above the black dotted curve corresponds to the WD regime where $\langle v_x \rangle_t=v_M$. At the lower edge of this region, the color map changes from blue (low $\langle v_x \rangle_t$) to red (high $\langle v_x \rangle_t$) as $U_M$ increases.  
 
As $v_M$ increases, the range of $U_M$ over which dragging occurs decreases and vanishes at $v_M=\Delta d/\hbar \pi=3.5$ mm s$^{-1}$, when $U_M^d=U_M^b$ (crossing of solid and dotted black curves in Fig. \ref{fig:4}).  At this $v_M$ value, the tilt of $E'(p_x)$ is large enough to ensure that the value of $U_M$ needed for the atom to traverse the first local minimum in $E'(p_x)$ is the same, or larger, as that required to traverse the first local maximum.  Therefore, when $U_M\approx U_M^d$, the dynamics change straight from the LD to the BO regime, i.e. without traversing the WD region.  

In the BO regime, we expect the atom to have its highest value of $\langle v_x \rangle_t$ when $U_M=U_M^d$ since here the atom spends most {\it time} in the region of the $E'(p_x)$ curve where the gradient of $E'(p_x)$, and thus $v_x'$, is positive.  This is confirmed by the color map in Fig. \ref{fig:4}, which reveals a narrow region of high $\langle v_x \rangle_t$ values (red), where $\langle v_x\rangle_t \sim v_M$,  just above the dotted black curve along which $U_M=U_M^d$.  

Also included in Fig. \ref{fig:4} are black and red dot-dashed curves showing, respectively, $U_M^{r_{min}}$ [Eq. (\ref{eq:UMrb})] and $U_M^{r_{max}}$ [Eq. (\ref{eq:UMrt})] for $r_{min}$ and $r_{max}$ = $1,2$ and $3$.  These curves show good correspondence with the positions of $\langle v_x \rangle_t$ peaks in the color map, which appear as blue / green or yellow stripes for $U_M\gtrsim 40$ peV = $0.390 E_R$.  

\section{Effect of initial position}
\label{sec:init_pos}

\begin{figure}%f1
 \centering
 \includegraphics*[width=1.\linewidth]{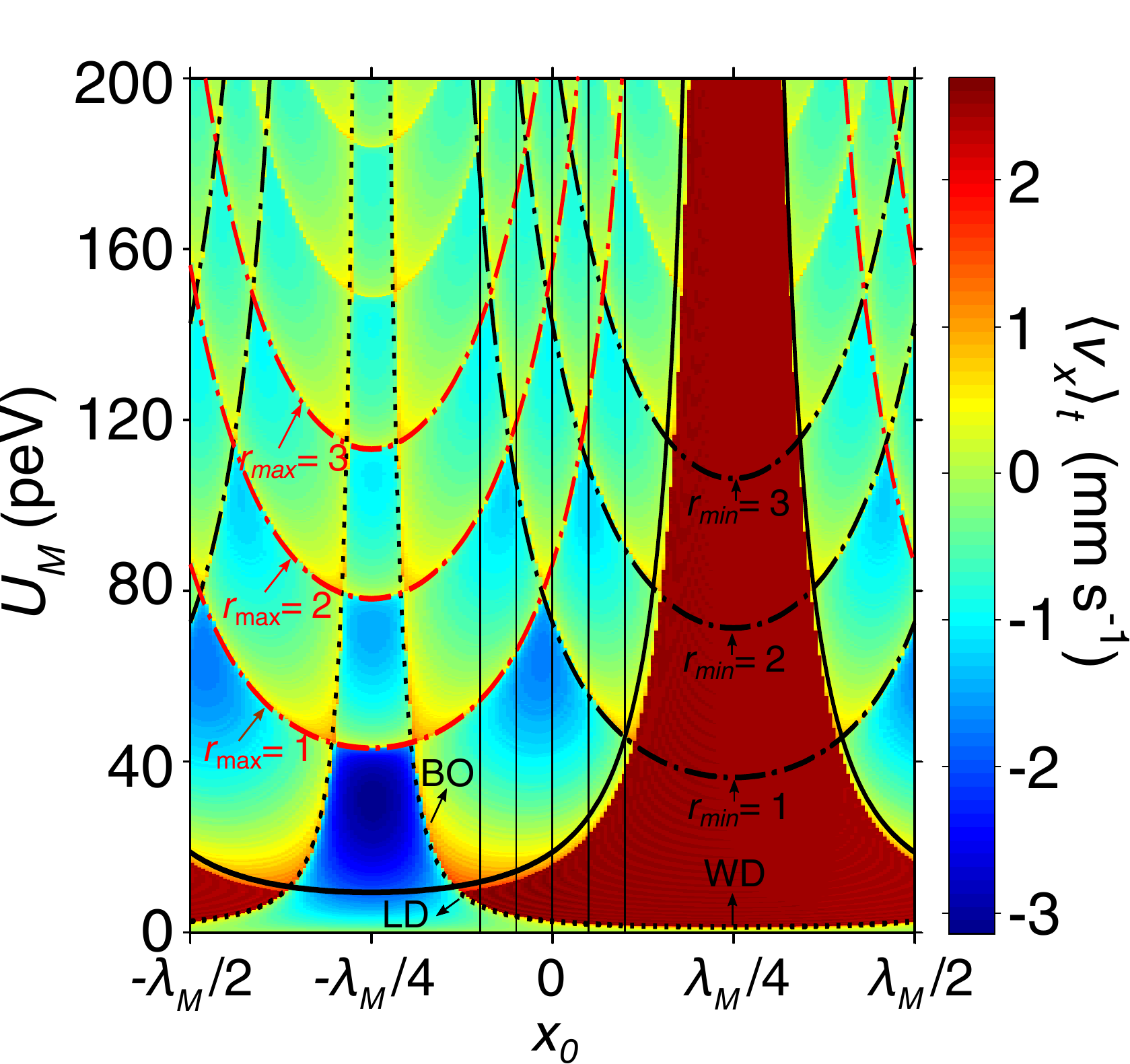}
  \caption{Color map (scale right) showing the variation of the average atom velocity with $x_0$ and $U_M$.  Black dotted curve shows the value of $U_M^d(x_0)$ representing the onset of dragged atom trajectories [Eq. (\ref{eq:Umd})].  Black solid curve marks the values of $U_M^b(x_0)$ [Eq. (\ref{eq:Umb})].  Black dot-dashed curves show $U_M^{r_{min}}(x_0)$ when $r_{min}=1,2$ and $3$ [Eq. (\ref{eq:UMrb})].  Red dot-dashed curves are the values of $U_M^{r_{max}}(x_0)$ where $r_{max}=1,2$ and $3$ [Eq. (\ref{eq:UMrt})].  For guidance, vertical black lines show the positions of the center of five wells in the SOL, located at $-2d$, $-d$, $0$, $d$ and $2d$. Labels LD, WD and BO mark regions of `Linear Dispersion', `Wave Dragging' and `Bloch Oscillations'.  
\label{fig:5}}
\end{figure}

Equations (\ref{eq:Umd}) and (\ref{eq:Umb}) indicate that the transitions between distinct dynamical regimes depend on the initial position of the atom $x(t=0)=x_0$. To illustrate this, Fig. \ref{fig:5} shows  a color map of $\langle v_x \rangle_t$ versus $x_0$ and $U_M$  when $v_M=2.5$ mm s$^{-1}$ and, as before, $\lambda_M=20d$. The color map has an intricate structure which, we now explain, is associated with the transition between the various dynamical regimes.

The solid curve in Fig. \ref{fig:2} corresponds to a vertical slice through the color map in Fig. \ref{fig:5}, when $x_0=0$, i.e. along the middle of the five vertical black lines. Increasing $U_M$ along this line, induces a transition from the LD into the WD regime on crossing the black dotted curve where $\langle v_x \rangle_t=v_M$.  The WD regime appears as the dark red region in the color map. With further increase of $U_M$, we enter the BO regime (on crossing the black solid curve), where the color map changes abruptly from red to yellow and the atom's velocity falls dramatically. Within the BO regime there is a series of resonances, which appear as yellow areas in Fig. \ref{fig:5} and correspond to those arrowed in Fig. \ref{fig:2}.  As explained in Section \ref{sec:semiclass} and Appendix \ref{app:2}, these resonances occur whenever $U_M$ coincides with a local extremum in $E'(p_x)$.

The variation of $U_M^d$ with $x_0$ [Eq. (\ref{eq:Umd})], shown by the dotted curve in Fig. \ref{fig:5}, agrees well with the transition from the LD to the WD regime seen in the color map.  As $x_0$ increases from $0$ to $\lambda_M/4$ the transition to the WD regime shifts slowly to lower $U_M^d$, which attains a minimum value of $1.3$ peV = $1.27 \times 10^{-2}E_R$ when $x_0\approx\lambda_M/4$.  For $x_0>\lambda_M/4$, $U_M^d$ increases slowly until $x_0=\lambda_M/2$, at which point the dynamics are {\it approximately} equivalent to those at $x_0=0$.  Conversely, decreasing $x_0$ below $0$ makes $U_M^d$ increase rapidly until $x_0\approx-3\lambda_M/16$ when the dotted curve in Fig. \ref{fig:5} diverges. As $x_0$ decreases further, we enter a regime in which there is no transition to the WD regime.  When $x_0\approx -5\lambda_M/16$, the dotted curve re-emerges and the value of $U_M^d$ decreases rapidly with decreasing $x_0$ until $x_0=-\lambda_M/2$, which is {\it exactly} equivalent to $x_0=\lambda_M/2$, meaning that the color maps on the left- and right-hand edges of Fig. \ref{fig:5} are identical.  

Equation (\ref{eq:Umd}) shows that as $v_M$ increases from 0, $U_M^d(x_0=0)$ also increases and, therefore, the range of $U_M$ values corresponding to the WD regime is reduced.  Figure \ref{fig:5} and Eq. (\ref{eq:Umd}) show, in addition, that as $x_0\rightarrow-\lambda_M/4$, $U_M^d(x_0)\rightarrow\infty$, meaning that for {\it any} given $v_M$ and $U_M$ the atom will {\it never} enter the WD regime.  We understand this in the effective dispersion curve picture by noting, from Eq. (\ref{eq:hamallt}), that when $x_0=-\lambda_M/4$, $H'=U_M$ and so $E'(p_x)$ can only take values between $0$ and $U_M$.  Therefore, the atom cannot traverse the local minimum in $E'(p_x)$ when %$E'(p_x)<0$ and 
$p_x\approx 0$ [arrowed in Fig. \ref{fig:3}(b)] and thus cannot enter the WD regime.  

The variation of $U_M^b$ with $x_0$ [Eq. (\ref{eq:Umb})], shown by the black solid curve in Fig. \ref{fig:5}, is in excellent agreement with the transition from the WD regime (red region in Fig. \ref{fig:5}) to the BO regime (above both the black dotted and solid curves in Fig. \ref{fig:5}).  As $x_0$ increases from 0, $U_M^b$ increases until $x_0\approx3\lambda_M/16$ where, for the range of $U_M$ shown, there is no transition to the BO regime above the red region.  Equation (\ref{eq:Umb}) shows that as $x_0\rightarrow \lambda_M/4$, $U_M^b \rightarrow \infty$, implying that trajectories with $x_0=\lambda_M/4$ are never able to Bloch oscillate for any $v_M$ or $U_M$.  Physically this is because when $x_0=\lambda_M/4$, according to Eq. (\ref{eq:hamallt}), $E'(p_x)$ only takes values between $-U_M$ and $0$, so the atom is unable to traverse the local maximum in $E'(p_x)$ [arrowed in Fig. \ref{fig:3}(d)] and thus cannot Bloch oscillate.  Figure \ref{fig:5} reveals that the Bloch oscillation regime reappears in the parameter space when $x_0\approx5\lambda_M/16$.  Thereafter, $U_M^b$ (right-hand segment of the black solid curve) decreases with increasing $x_0$ until $x_0=\lambda_M/2$, which is equivalent to $x_0=-\lambda_M/2$ (color maps are identical on left- and right-hand edges of Fig. \ref{fig:5}).  

As $x_0$ decreases from 0, the transition from the WD to the BO regime occurs at slowly decreasing $U_M$ values until $U_M^d(x_0)=U_M^b(x_0)$ (where the black dotted and solid curves cross in Fig. \ref{fig:5}).   As $x_0$ decreases further, the two curves cross again when $x_0\approx -3\lambda_M/8$.  For $x_0$ values between these two crossing points, there is no wave dragging regime and, for given $x_0$, increasing $U_M$ induces a transition straight from the LD to the BO regime upon crossing the black solid curve. As $x_0$ approaches $-\lambda_M/4$ then, for any given $U_M$, the atom cannot traverse the local minimum in $E'(p_x)$ [arrowed in Fig. \ref{fig:3}(b)] and thus also does not reach the first local maximum in $E'(p_x)$ [arrowed in Fig. \ref{fig:3}(d)].  Therefore, the atom will only change direction when it reaches the peak in $E'(p_x)$ occurring when $p_x\approx-\hbar\pi/d$, resulting in the dramatic change from negative to positive velocity [transition from blue to yellow regions in Fig. \ref{fig:5}] when $U_M\approx45$ peV = $0.438E_R$.  

As discussed previously, the strong resonant features in $\langle v_x \rangle_t$ versus $U_M$ and $x_0$ in Fig. \ref{fig:5} occur when the atom accesses new peaks in $E'(p_x)$ [see Eqs. (\ref{eq:UMrb}) and (\ref{eq:UMrt})].  The variation of $U_M^{r_{min}}$ and $U_M^{r_{max}}$ with $x_0$ shown by the black and red dot-dashed curves respectively in Fig. \ref{fig:5} for (bottom to top) $r_{min}$ and $r_{max}=1,2$ and $3$ agree well with the yellow regions of the color map where $\langle v_x \rangle_t$ increases abruptly.  Note that near these resonances, the color map changes abruptly between yellow and blue indicating that $\langle v_x \rangle_t$ changes suddenly from positive to negative: i.e small changes in $x_0$ or $U_M$ switch the atom's direction of travel. 

\section{Quantum calculations}
\label{sec:quantum}

In this section, we investigate the full quantum-mechanical description of the semiclassical dynamical regimes revealed in Sections \ref{sec:semiclass}-\ref{sec:init_pos}. The quantum-mechanical Hamiltonian for the system is

\begin{equation}
 \hat{\mathcal{H}} = -\frac{\hbar^2}{2m_a} \frac{\partial ^2}{\partial x^2}  + V_{OL}(x) + V_M(x,t),
\label{eq:OLHamiltonian}
\end{equation}

\noindent
where $m_a$ is the mass of a single $^{23}$Na atom. We solved the corresponding time-dependent Schr\"odinger equation
 
\begin{equation}
 i \hbar \frac{\partial \psi(x)}{\partial t} = \hat{\mathcal{H}}(x,t) \psi(x),
\label{eq:timedependschro}
\end{equation}

\noindent
numerically using the Crank-Nicolson method \cite{NUMERICAL}.  Note that we consider the mean-field interaction to be negligible compared with the external potential.  This can be realised in experiment by using an applied magnetic field to tune the Feshbach resonance so that the inter-atomic scattering length is zero.  By only considering a non-interacting atom cloud we can present a full and detailed description of the fundamental dynamical phenomena exhibited by the system.  If atom interactions are included we expect more complex dynamics which would deserve further study and analysis.

\begin{figure}%f1
  \centering
  \includegraphics*[width=1.\linewidth]{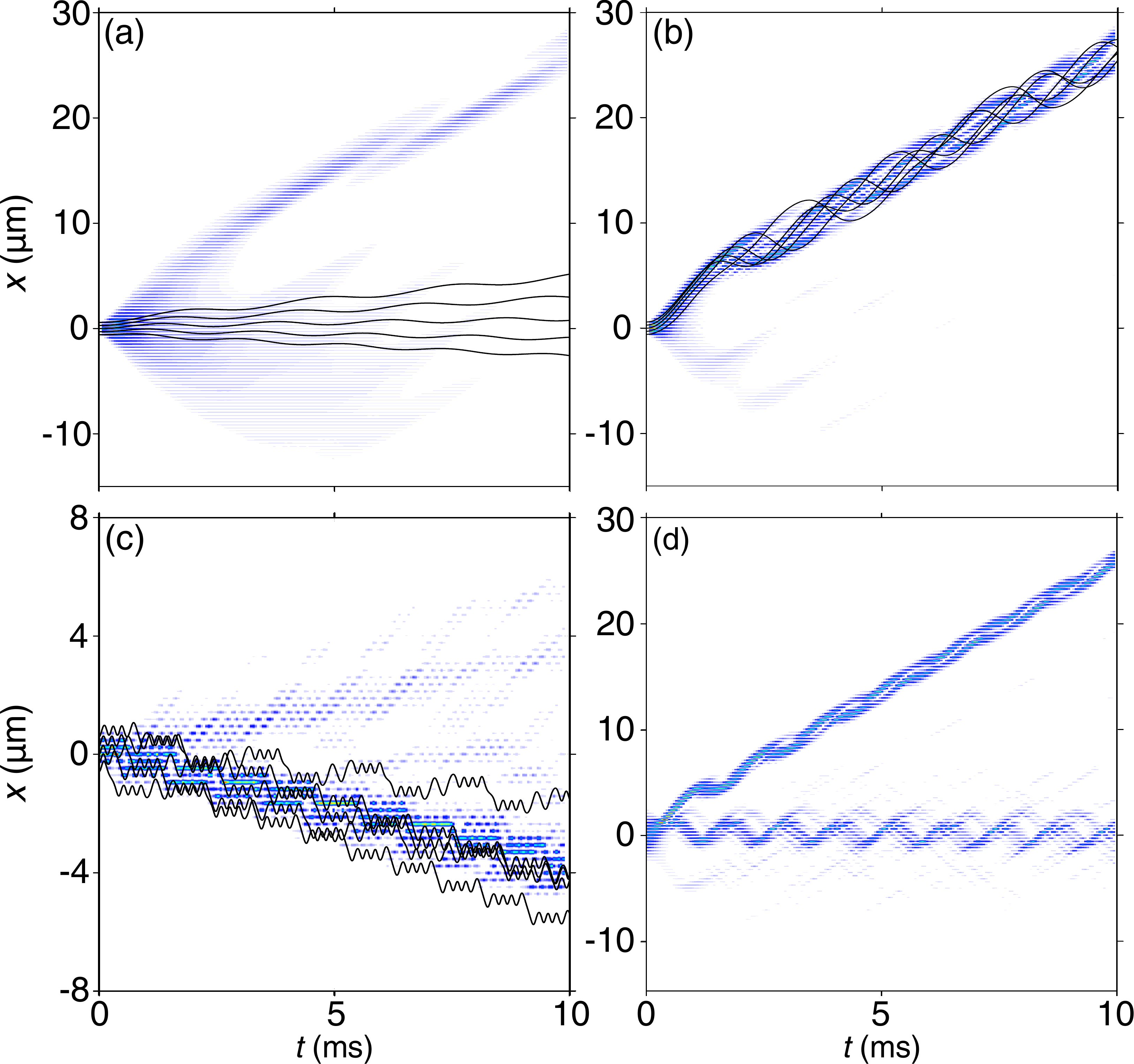}
   \caption{Gray-scale maps [color online: white = 0,  blue = high] showing the evolution of the atom density for: (a) $U_M=1$ peV = $9.74 \times 10^{-3}E_R$ and $f_x=2d$ (LD regime); (b) $U_M=10$ peV = $9.74 \times 10^{-2}E_R$ and $f_x=2d$ (WD regime); (c) $U_M=150$ peV = $1.46E_R$ and $f_x=2d$ (BO regime); (d) $U_M=25$ peV = $0.243E_R$, $f_x=8d$ which encompasses both the Wave Dragging and Bloch Oscillation regimes. See text for further details. Solid curves in (a)-(c) show corresponding semiclassical trajectories with, from bottom to top, $x_0=-2d$, $-d$, $0$, $d$ and $2d$.  Movies showing the time evolution of the wavepackets are shown in the Supplemental Material \cite{suppmat}. \label{fig:6}}

\end{figure}

First, we found the form of the initial wavepacket by evolving Eq. (\ref{eq:timedependschro}) in imaginary time, using an initial Gaussian wavefunction with a full width half maximum value, $f_x=2d$, to produce a state similar to the ground state of an atom cloud held in the SOL by a harmonic trap.  To study the atomic dynamics, we then integrated Eq. (\ref{eq:timedependschro}) numerically, using this wavepacket as the initial state, which, when $f_x=2d$, extends across $\sim 5$ wells of the SOL. Figure \ref{fig:6} shows the time-evolution of the wavepacket in each of the three distinct dynamical regimes corresponding to the semiclassical trajectories shown in Fig. \ref{fig:3} ($v_M=2.5$ mm s$^{-1}$, $\lambda_M = 20 d$) (see also Supplemental Information \cite{suppmat}).  The central black curve in Fig. \ref{fig:6}(a) corresponds to the semiclassical trajectory in the LD regime ($U_M=1$ peV, $x_0=0$) shown in Fig. \ref{fig:3}(a) and Supplemental Information, Movie 1 \cite{suppmat}. Also shown are semiclassical trajectories calculated for (from bottom to top) $x_0=-2d,-d, 0, d$ and $2d$, i.e. for orbits starting at the centers of the 5 wells that are spanned by the initial wavefunction.  In this case, most of the wavepacket, shown by the gray-scale map in Fig. \ref{fig:6}(a), does not follow the semiclassical trajectories.  Instead, it spreads rapidly with only a small fraction of the wavefunction being dragged through the lattice.  

From Fig. \ref{fig:5}, at first sight we might expect that the wavepacket lies entirely in the LD regime because, at $U_M=1$ peV = $9.74 \times 10^{-3}E_R$, the vertical lines marking the centers of the potential wells spanned by the wavepacket all lie in the light green region below the black dotted curve, i.e. fully within the LD regime.  It is clear from Fig. \ref{fig:6}(a), however, that
the wavepacket does not follow a LD semiclassical trajectory.  This is because for such a low $U_M$ value the spreading of the wavepacket, resulting from position-momentum uncertainty, dominates the dynamics obtained from the semiclassical analysis. From the uncertainty principle, $f_x\Delta p_x \approx \hbar$, where $\Delta p_x$ is the spread of the momentum of the wavepacket, and the corresponding zero point energy of the wavepacket is $E_{zp}=\Delta p_x^2 / 2 m_a$.  When $f_x=2d$, $E_{zp}\approx2.5$ peV = $2.43 \times 10^{-2}E_R$.  This energy is large enough to allow part of the wavepacket to enter the wave dragging regime, as observed in Fig. \ref{fig:6}(a). This zero point energy determines a lower limit of $U_M$ below which the
semiclassical analysis becomes invalid due to being dominated by wavepacket expansion.

Figure \ref{fig:6}(b) and Supplemental Information, Movie 2 \cite{suppmat}, shows the evolution of the wavepacket when $U_M=10$ peV = $9.74 \times 10^{-2}E_R$, corresponding to the semiclassical trajectory in the WD regime shown in Fig. \ref{fig:3}(c).  The striking similarity between the semiclassical orbits and the atom density evolution reveals that, in this regime, the wavepacket is dragged through the SOL by the MOL. 
The reason for this is clear from Fig. \ref{fig:5}, which reveals that for $U_M=10$ peV the full spread of the wave function (between the left- and right-most dashed vertical lines in Fig. \ref{fig:5}) lies within the red wave dragging regime.  For this $U_M$ value, the moving lattice is strong enough to dominate any spreading caused by the zero point energy.

Figure \ref{fig:6}(c) and Supplemental Information, Movie 3 \cite{suppmat}, shows the wavepacket evolution for $U_M=150$ peV = $1.46E_R$, corresponding to the BO regime [e.g. the orbit in Fig. \ref{fig:3}(e)]. The five semiclassical trajectories shown in Fig. \ref{fig:6}(c) broadly encompass the evolution of the negative-going part of the wavepacket. One might expect that the energy band would break for this large value of $U_M$, causing the wavepacket to no longer track the semiclassical dynamics.  However, we note that the relatively long wavelength $\lambda_M=20d$ of the moving lattice ensures that the energy drop across a single well within the SOL is $\approx45$ peV = $0.438E_R$, which is small enough to keep the band intact.  Note also for high values of $U_M$ the approximation that the atom will remain in the lowest energy miniband becomes invalid, we estimate this upper limit on $U_M$ by assuming that interband tunnelling will occur when the amplitude of the moving wave is approximately equal to the energy gap between the first and second minibands. The two minibands calculated for the stationary optical lattice are $\approx 250$ peV apart [see Fig. \ref{fig:1}(a)].  Therefore we expect interband tunnelling to occur when $U_M\gtrsim250$ peV.  It appears, from our numerical simulations, that for $U_M\gtrsim200$ peV = $1.95E_R$,  the wavepacket dynamics start to deviate from the semiclassical paths due to band breakdown.  

The sensitive dependence of the atom dynamics to the initial position, $x_0$, shown in Fig. \ref{fig:5}, has an interesting consequence. Namely,
if the atom cloud is large enough, it can simultaneously span more than one dynamical regime.  For example, in Fig. \ref{fig:6}(d) and Supplemental Information, Movie 4 \cite{suppmat}, we show the wavepacket evolution for $U_M=25$ peV = $0.243E_R$, with an initial atom cloud width $f_x=8d$ and center of mass position $x_0=1.5d$.  The figure shows that the atom cloud splits, with the lower part performing Bloch oscillations around $x=0$ and the rest being dragged through the SOL by the MOL.  The reason is, as can be seen from Fig. \ref{fig:5}, that the upper part of the initial atom cloud with $x\gtrsim1.5d$ lies in the wave dragging regime whilst the lower part with $x\lesssim1.5d$ is in the Bloch oscillation regime.  Our semiclassical analysis suggests that approximately half the cloud is in the wave dragging regime.  This estimate agrees well with the numerical evolution of the wavefunction, which reveals that after 10 ms, the fraction of the cloud in the wave dragging regime is $0.495$. This result suggests that information about the shape of the initial atom cloud could be obtained by measuring the number of atoms, $N_U$, that end up in the upper (dragged) part of the split cloud, as a function of $x_0$.  For example, as $x_0$ increases so that more of the initial wavepacket lies within the red WD regime in Fig. \ref{fig:5}, i.e. so that the initial wavepacket is shifted across the black solid curve, more atoms will end-up in the upper part of the split cloud.  More quantitatively, we expect that $N_U(x_0)\approx\int^\infty_{x_S(U_M)}\left|\Psi(x_0,t=0)\right|^2 dx$ where $x_S$ satisfies $U^b_M(x_S)=U_M$, i.e. marks the position of the solid curve in Fig. \ref{fig:5} for given $U_M$.

\section{Conclusions}
\label{sec:conc}

We have shown that a periodic optical potential moving through a stationary optical lattice can control the atom orbits in real and momentum space, and induce transitions between three distinct dynamical regimes.  In a semiclassical picture, the atom can be confined to a small region of the stationary optical lattice dispersion curve, dragged through the stationary optical lattice by the moving wave, or perform Bloch-like oscillations.  The crossover between these orbital types creates a rich pattern of transport regimes, with multiple resonances that cause abrupt changes in the magnitude and sign of the atom's average velocity.  These resonances occur whenever there is an abrupt change in the number of Brillouin zones that the atom can access. Since the form of the atom trajectories is predictable, and can be tailored by small changes in the MOP parameters, the system provides a mechanism for moving atom clouds between well-defined points in stationary optical lattices. It may also provide a sensitive method to split atom clouds, with potential uses in atom interferometry and in mapping the initial atom cloud dynamically rather than via optical imaging.  The dynamics described in this paper are similar to those expected for electrons in acoustically-driven superlattices \cite{GRE2010}.  Consequently, the cold-atom system that we have studied also provides a quantum simulator for acoustically-driven semiconductor devices in which the electron orbits are much harder to observe directly.  

\appendix
\section{Detailed description of atom motion in the Linear Dispersion regime}
\label{app:1}
In the LD regime, initially, for $t=0$ and $x'=0$, the force applied to the atom by the moving optical lattice, is maximal, which makes $p_x$ increase until $E'(p_x)=-U_M/2$ [lower filled circle in Fig. \ref{fig:3}(b)]. At this point, the atom's velocity in the {\it rest frame}, $v_x=\partial E'(p_x) \partial p_x+v_M$, attains its maximum value for the specified initial condition.  The potential energy $V_M(x')=U_M$ (to keep $H'$ constant) is also maximal, and so the force is zero in the moving frame.  However, in the moving frame $x'$ continues to decrease, which eventually makes the force negative ($-dV_M(x')/dx'<0$), so accelerating the atom in the negative $p_x$ direction.  

The atom continues to move in the negative $p_x$ direction until $E'(p_x)=U_M/2$ [upper filled circle in Fig. \ref{fig:3}(b)] where the atom's velocity in the rest frame is minimal.  The atom then returns to its initial condition with $p_0=0$ and the cycle repeats with the atom traversing successive regions where $V_M<U_M/2$ [white regions in Fig. \ref{fig:3}(a)] and $V_M>U_M/2$ [gray regions in Fig. \ref{fig:3}(a)].  Since the region of the dispersion curve accessed by the atom is not completely linear, the velocities at the extremities of the orbit are slightly different, causing the atom to have non-zero average velocity. 

\section{Detailed description of atom motion in the Bloch Oscillation regime}
\label{app:2}

In the BO regime, the atom can access $E'(p_x)$ across several Brillouin zones, which makes $\langle v_x\rangle_t$ vary in a complex way with increasing $U_M$.  For $U_M=150$ peV = $1.46E_R$, the lowest possible value of $E'(p_x)$ [lower dashed line in Fig. \ref{fig:3}(f)] occurs just below a local minimum in the curve [right-hand arrow in Fig. \ref{fig:3}(f)].  The velocity of the atom in the moving frame, $\partial E'(p_x)/\partial p_x$, is negative at the right-hand filled circle, indicating that the atom is moving in the negative $x'$ direction. However, since $V_M(x')$ is maximal at this point, the moving lattice exerts little force on the atom. Consequently, $p_x$ decreases only slowly and, as a result, the atom also lingers in the region of $E'(p_x)$ where $v_x'$ (and thus $v_x$) is \emph{positive} [within the region marked by the right-hand bracket in Fig. \ref{fig:3}(f)].  Overall, as the atom moves away from the right-hand filled circle in Fig. \ref{fig:3}(f), it spends more time with positive than negative velocity causing the atom to jump {\it forwards} along the $x-$axis [arrows labelled `F' in Fig. \ref{fig:3}(e)]. 

Conversely, the maximum attainable value of $E'(p_x)$ [upper dashed line in Fig. \ref{fig:3}(f)] occurs just below the local maximum in $E'(p_x)$, marked by the left-hand arrow in Fig. \ref{fig:3}(f), where $V_M$ is maximal.  Since the force on the atom is low in this part of the $E'(p_x)$ curve, the atom remains in the negative velocity region of $E'(p_x)$ [left-hand bracket in Fig. \ref{fig:3}(f)] for a long time. This makes the atom take a large jump {\it backwards} along the $x-$axis [arrows labelled `B' in Fig. \ref{fig:3}(e)].  Figure \ref{fig:3}(e) reveals that for $U_M=150$ peV, the atom jumps backwards further than it jumps forwards, so giving the atom an overall negative average velocity. 

If $U_M$ has a value that just exceeds a local maximum or falls just below a local minimum in the $E'(p_x)$ curve, then the time that the atom spends in the region of the $E'(p_x)$ curve, where $dE'(p_x)/dp_x=v_x'>0$, is maximised.  Consequently, the magnitude of the forward jump in the orbit, and therefore  $\langle v_x \rangle_t = \langle v_x' \rangle_t + v_M$, is maximised. This phenomenon produces the series of peaks observed in the $\langle v_x \rangle_t$ versus $U_M$ curve [see arrowed peaks in Fig. \ref{fig:2}], the positions of which are discussed in the main text.

\end{document}